# Symmetry breaking and mechanical filter make a pseudo-gimbal-less two-dimensional MEMS scanning mirror with multiple scanning modes

Weimin Wang

Key Laboratory of Optoelectronic Technology and Systems, Ministry of Education, Defense Key Disciplines Laboratory of Novel Micro-Nano Devices and System Technology, College of Optoelectronic Engineering, Chongqing University, Chongqing 400044, China, weiminwang@qq.com

## Abstract

Miniaturized two-dimensional scanning mirror based on microelectromechanical systems (MEMS) technology has great potential in automotive industry, consumer electronics, and biomedicine, etc. Due to its high frequency and large angle, resonant scanning is the mainstream in all MEMS actuation mechanisms, such as harmonic resonant electromagnetic scanner and parametric resonant electrostatic scanner. Although electrostatic scanner has the advantages of low power consumption and IC process compatibility, some shortcomings of parametric resonance, including double frequency of driver electronics and additional feedback control or frequency stabilization system, limit its further application. The symmetry of coplanar electrostatic comb actuator is broken in this paper, and harmonic resonant electrostatic scanner with excellent performance is realized. Further, through adopting mechanical filter, two-dimensional scanning can be achieved through one set of actuators, which avoids the problem that two sets (each for one dimension) of electrostatic actuators must be insulated each other through complicated and expensive processes. A two-dimensional MEMS scanner based on symmetry breaking and mechanical filter was proposed and demonstrated. Multiple scanning modes can be achieved through selective control of a set of four identical actuators.

# Introduction

In recent years, due to huge demands in the fields of Light Detection and Ranging (LiDAR) [1-6], wearable near-to-eye display [7,8], and vehicle Head Up Display (HUD) [9], among others, microelectromechanical systems (MEMS) scanning mirror technology attracts extensive attention. Many scanners based on various MEMS actuators have been developed, among which the electrostatic comb drive actuated devices have the advantages of low power consumption, IC process compatibility, and compactness [10,11]. MEMS scanning mirror has many performance parameters, such as aperture, angle, frequency, power consumption, costs and so on. In addition to some general parameters, different application fields are concerned with different parameters.

The development trend of MEMS scanning mirror includes larger angle, higher frequency, monolithic integrated two-dimensional scanning, etc. In order to increase the scanning angle and frequency, resonance mechanism is widely used. However, most of the current comb driven resonant scanners are based on parametric resonance [12-18] which brings several problems. First, the driving signal must be at the twice the resonant frequency. Thus, as the frequency of the scanning mirror increases, the frequency of power supply increases more rapidly, which will have a serious impact on the cost and volume of the scanning mirror. Second, for parametric resonance, a frequency shift of 10 ppb (parts per billion) at the resonant peak stops the oscillation due to the presence of an unstable region[19]. As a result, additional feedback control or frequency stabilization system is needed. Third, a change in the driving voltage will cause the same phenomenon as the frequency shift. Under a stable parametric vibration, lowering the voltage does not reduce the scanning angle, but is very likely to stop the vibration. In other words, the mirror has a minimum switching voltage. This puts forward higher demand on power supply and makes angle adjustment more complicated. Aiming at the above problems existing in scanning mirror based on parametric resonance, a non-parametric resonance (traditional harmonic resonance) structure is proposed in this paper, and the scanning performance as parametric resonance is achieved.

Two-dimensional (2D) MEMS scanners are more powerful than one-dimensional scanners, but this comes at the cost of greater development difficulty. Gimbal-less and gimbaled scanner can both achieve 2D scanning trajectory. Gimbal-less scanners generally adopted a completely symmetrical structure and three [20] or four [21,22] identical actuators, resulting in the same resonant frequency in two axis. Therefore, gimbal-less scanners are not suitable for resonant scanners because two orthogonal vibrations of identical scan frequency only produce circular or linear trajectory. The gimbaled scanner has an inner axis and an outer axis (also known as fast axis and slow axis) which can produce different resonant frequencies.

As shown in Fig. 1a, the two axis of the gimbaled scanner each have their own actuators. In order to insulate the two electrodes of the inner actuators located on the frame from each other, a high level of fabrication processes, for example, the etching and refilling of high aspect ratio insulation trench [23,24], the combination of the surface- and the bulk- micromachining techniques[25], the suspended substrate[26], flip-chip bonding and fusion bonding of one SOI wafer and three silicon wafers[27], and so on, is required.

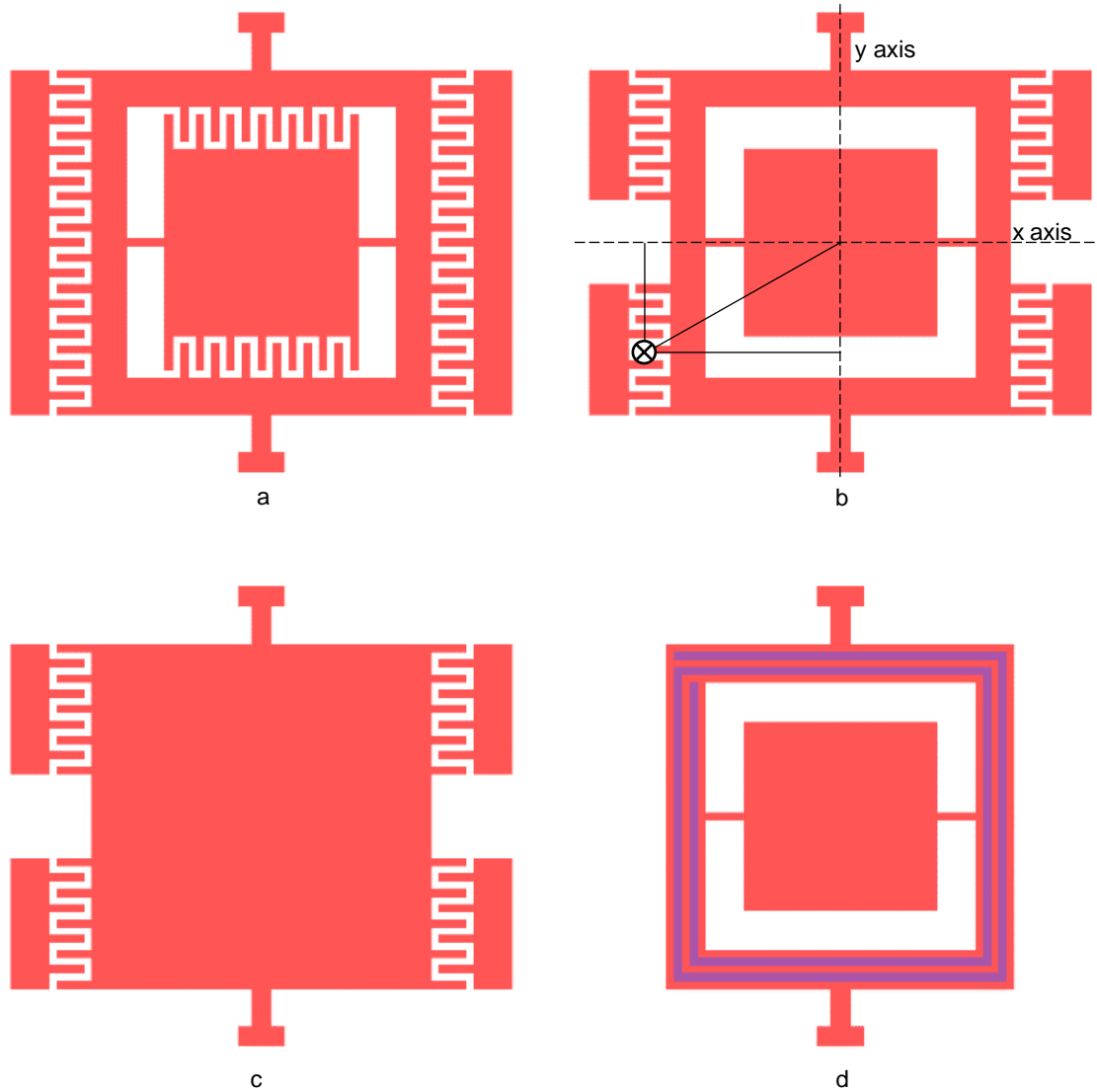

Fig. 1 Structures of several MEMS scanner. a Conventional gimbaled electrostatic comb actuated scanner. b Proposed gimbaled electrostatic comb actuated scanner. c Proposed gimbal-less electrostatic comb actuated scanner. d Conventional gimbaled electromagnetic scanner.

To overcome respective shortcomings of gimbal-less and gimbaled structures, a pseudo-gimbal-less 2D electrostatic scanner with four identical comb-drive actuators is proposed. As shown in Fig. 1b and c, the scanner can employ gimbal or gimbal-less frame. All actuators are placed outside the frame, hence the structure does not need insulation and can be fabricated by the most conventional SOI bulk micromachining. When any actuator is driven, the driving torque can be decomposed to two perpendicular torques, around fast and slow scan directions respectively, as shown in

Fig. 1b.

Based on the mechanical filter principle which has been used in electromagnetic scanners [28,29], a superimposed torque comprising high frequency and low frequency components is applied. Both components are decomposed to two torques about fast and slow axis, but high frequency component about slow axis and low frequency component about fast slow cannot produce any response, known as high cut (low pass) and low cut (high pass) filtering. At last the movement of the scanner is a linear superposition of the response of fast axis to the high frequency torque and the response of slow axis to the low frequency torque.

In electromagnetic actuators, due to the linear relation between driving voltage and torque, one actuator with superimposed voltage is enough (Fig. 1d). However, voltage-torque square law for electrostatic actuators makes the desired torque cannot be obtained by a superimposed voltage. At least two actuators, each with the voltage corresponding to one of the frequency components, are needed.

## Results

**Design.** Figure 2 is the layout of the proposed scanner. To demonstrate the detail of the structure clearly, only one fixed comb is depicted. Structural parameters are listed in Table 1.

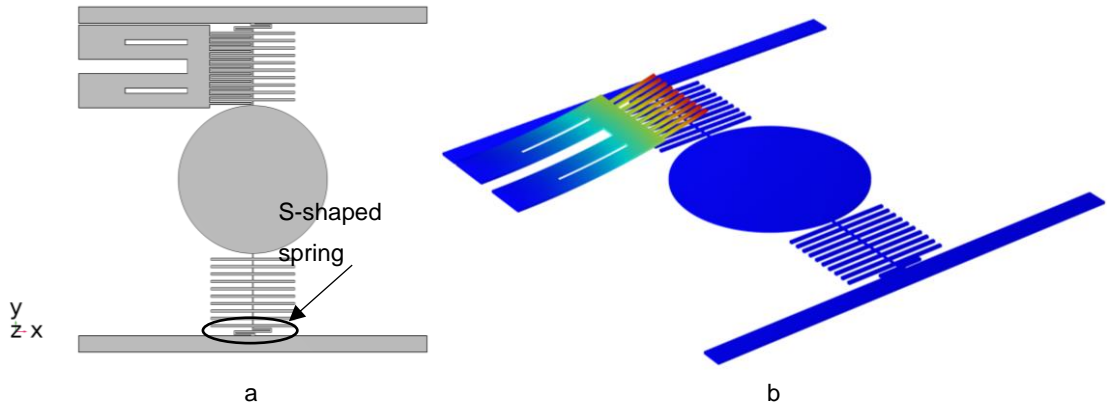

Fig. 2 Proposed electrostatic MEMS scanning mirror (only one fixed comb is depicted). a Vertical view. b Oblique drawing. The tip of the fixed comb finger is bent upward due to the residual stress

Table 1 Design parameters of the scanner

| Parameter | Value |
| --- | --- |
| Diameter of scanner | 1.8 mm |
| Thickness of scanner | 25 μm |
| Length of finger | 500 μm |
| Width of finger | 30 μm |
| Lateral spacing of finger | 15 μm |
| Offset length of finger | 15 μm |
| Number of moving fingers | 40 |
| Length of beams | 910 μm |
| Width of beams | 10 μm |
| Length of S-shaped spring | 450 μm |
| Width of S-shaped spring | 20 μm |

The structure is gimbal-less, similar to Fig. 1c. Therefore, the fixed end of the beam is designed to be S-shaped to decrease the torsional stiffness around the x-axis. Figure 3 is the simulated resonant torsional motion around the y-axis and x-axis of the device. The resonant frequencies are 1031 Hz (slow axis) and 6010 Hz (fast axis), respectively.

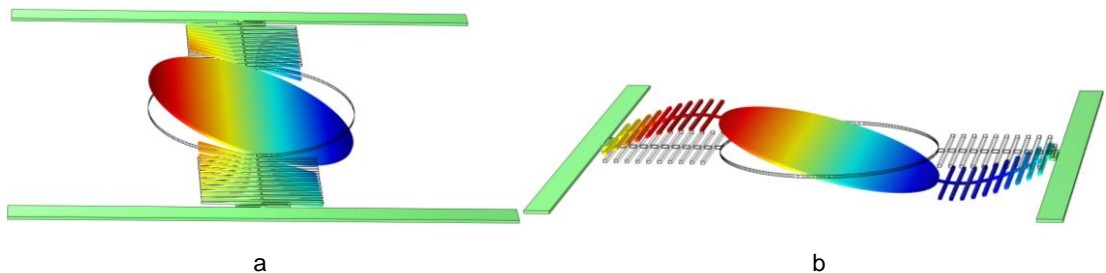

Fig. 3 Simulated resonant modes. a Around y-axis. b Around x-axis.

To avoid the shortcoming of parametric resonance, residual stress is used to induce the bending of the tip of the fixed comb finger. Thus, the coplanar comb actuator changed to non-planar comb actuator, as shown in Fig. 2b. Its principle can be explained physically and mathematically. Figure 4a show the cross-sectional view of the coplanar comb actuator. A voltage is applied between the fixed and moving comb. (i) and (iii) are the equilibrium position of the scanner vibration (in mechanical field), (ii) and (iv) are the positive and negative maximum. It's a complete cycle from (i) to (iv). However, in an electrostatic field, due to the structural symmetry with the horizontal and vertical dot dash line, the electric field distributions of (i) and (iii) are the same and the electric field distributions of (ii) and (iv) are the same, too. As a result, from (i) to (iv) is two cycles. So, the frequency of the electric field is twice that of the vibration, which is typical characteristic of a parametric excitation system.

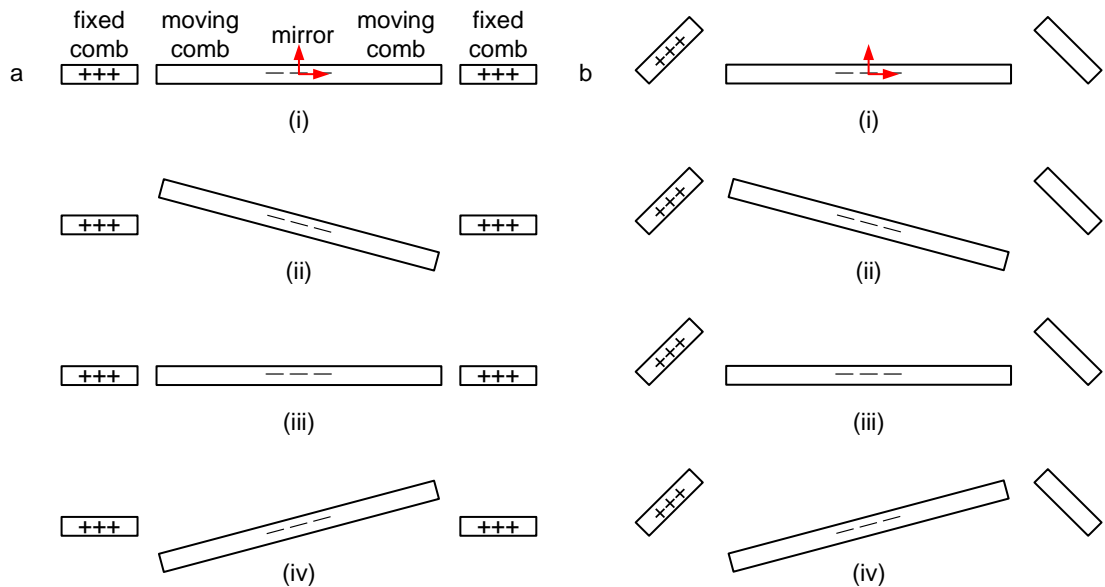

Fig. 4 A complete scanning cycle of coplanar and non-planar comb actuated scanner. a Coplanar scanner. b Non-planar scanner.

The non-planar comb can break the two symmetries, as shown in Figure 4(b). First, the curled fixed comb breaks the symmetry with horizontal dot dash line. Second, the voltage is applied only to the moving comb and left fixed comb. As a result, the symmetry with vertical dot dash line is also broken. Specifically, the electric field distribution in present (ii) and (iv) is different. It's just one cycle from (i) to (iv). Thus, the frequency of the driving voltage is the same as that of the resonance.

From a mathematical point of view, parametrically excited comb scanner can be modelled as [30]

$$I_m \frac{d^2\theta}{dt^2} + b\frac{d\theta}{dt} + k\theta = M(\theta) \qquad (1)$$

where $I_m$ is the moment of inertia, $\theta$ is the scanning angle, $b$ is the air damping, $k$ is the torsional stiffness, and $M$ is the electrostatic torque. It can be expressed as

$$M(\theta) = \frac{dE}{d\theta} = \frac{1}{2}V(t)^2 \frac{dC}{d\theta} \qquad (2)$$

where $E$ is the net energy, $V(t)$ is the periodic driving voltage, and $C$ is the total capacitance. For coplanar comb actuated scanner, A typical $dC/d\theta$ curve is shown in Figure 5. Though different comb shape will produce different curve, it must be odd function due to symmetry mentioned above.

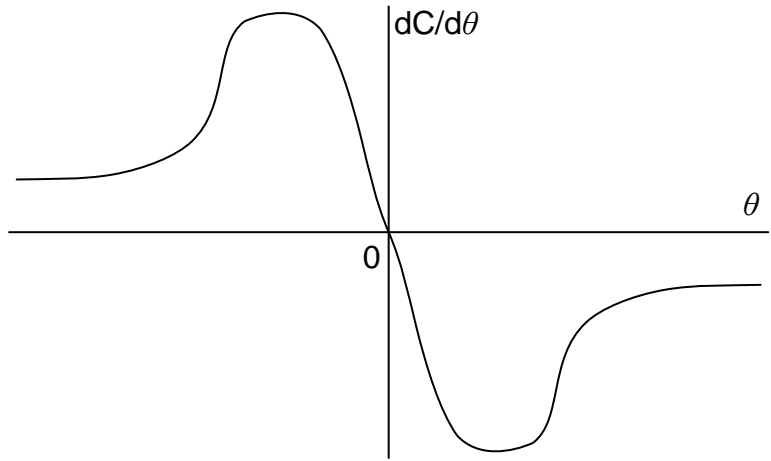

Fig. 5 Schematic diagram of the dC/d$\theta$ function of the coplanar comb actuator.

So $dC/d\theta$ can be expanded as odd power series

$$\frac{dC}{d\theta} = A\theta + B\theta^3 + \cdots\cdots \tag{3}$$

Substituting power series into the above differential equation, it can be simplified to a Mathieu equation with odd power nonlinear term [30]. Its solution shows that fundamental resonance is located twice the mechanical resonance frequency of the structure.

In the proposed device, due to the breaking of the symmetry, the *dC/dθ* curve is no longer an odd function, and, the resonant frequency is no longer twice the mechanical resonant frequency because the corresponding differential equation cannot be simplified to a Mathieu equation with odd power nonlinear term.

It is worth mentioning that, in Fig. 4b, if the fixed combs of the non-planar comb at both sides are driven together, the symmetry with vertical dot dash line is still satisfied and parametric resonance will be excited.

**Fabrication.** A prototype of the proposed device was fabricated by a commercially available SOI (Silicon on Insulator) micromachining process, SOIMUMPs provided by MEMSCAP Inc [31]. It is a simple 4-mask level SOI patterning and etch process. A photograph of the prototype is shown in Fig. 6a. Two metal layers (Pad Metal and Blanket Metal) with large residual stress were deposited on the surface of the four fixed combs. Three-dimensional profile was measured by a 3D white light profiler and shown in Fig. 6b. The height difference between the moving and the fixed comb is about 14 μm. The four actuators are numbered I, II, III, IV, as shown in Fig. 6a.

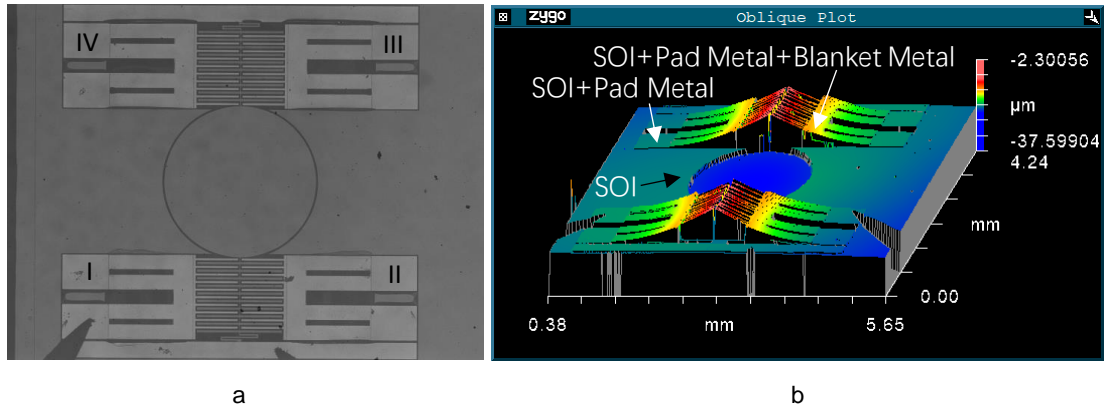

Fig. 6 Fabricated device. a Micrograph. b Measured 3D profile.

If only one actuator (I, II, III, or IV) is driven, according to different driving frequencies, the harmonic resonance around x axis or y axis can be selectively excited. If actuators I and II are driven together, due to the fact that their position is symmetric with respect to the y axis and nonsymmetric with respect to the x axis, the harmonic resonance around x axis or parametric resonance around y axis can be excited. The driving of actuators I and IV is just the opposite. Table 2 list the several driving modes.

Table 2 Multiple scanning modes

| Driving actuators | Mode |
| --- | --- |
| I/II/III/IV | harmonic resonance around x axis/harmonic resonance around y axis |
| I & II/III & IV | harmonic resonance or quasistatic torsion around x axis/parametric resonance around y axis |
| I & IV/II & III | harmonic resonance or quasistatic torsion around y axis/parametric resonance around x axis |

**Y axis (slow axis) testing.** A square wave of 20 V voltage was applied between the actuator I and the mirror, and its frequency was swept around the simulated slow axis resonant frequency. The oscillation angle of the mirror around the slow axis was measured by the Dynamic Metrology Module (DMM) of the profilometer, which freeze-frame the high-speed oscillation through a strobe light source with the same frequency, as shown in Fig. 7a. The measured angle is shown in Fig. 7c as the red curve.

As mentioned earlier, actuator I is driven singly will excite harmonic resonance and the measured resonant frequency is 990 Hz, very close to the simulation value of 1031 Hz.

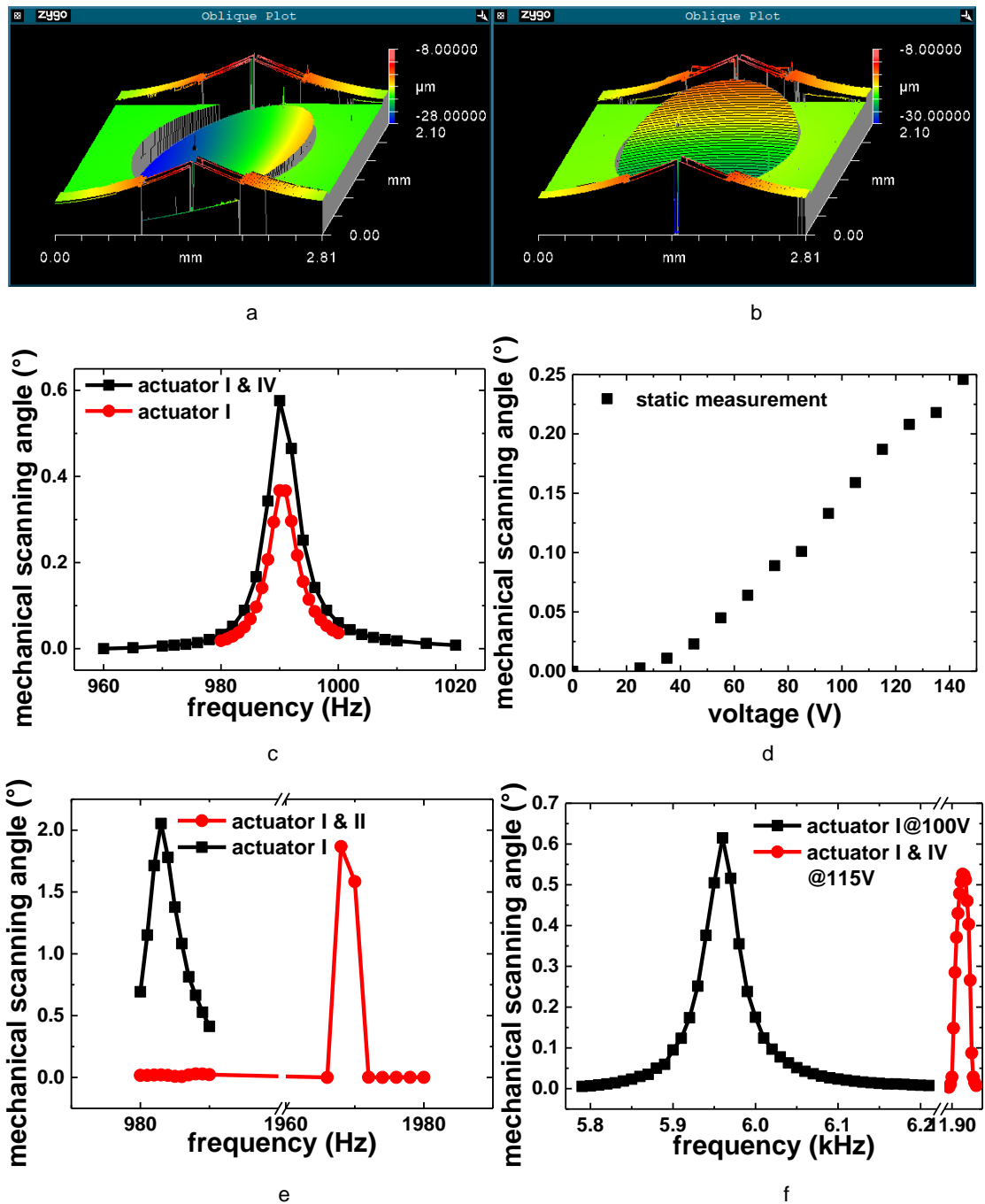

Fig. 7 Experimental results. a Scanning profile around y-axis. b Scanning profile around x-axis. c One and two actuators actuated harmonic resonance around y-axis. d Quasistatic scanning around y-axis. e Harmonic resonance and parametric resonance around y-axis. f Harmonic resonance and parametric resonance around x-axis.

Next, the same voltage was simultaneously applied to actuators I and IV, due to the fact that they are located on the same side of the y-axis. The results are plotted in Fig. 7c as the black curve. A maximum mechanical scanning angle of ±0.58° is achieved at frequency of 990Hz. Although the actuator has doubled, the angle has not doubled.

Actuators I and II are located on both sides of the y-axis, thus can be used to excite the parametric resonance. On the other hand, for parametric resonance, the minimum turn-on voltage must be satisfied. Through many experiments, it was found that the minimum turn-on voltage is 50 V. A 50-V square wave was applied between actuators I, II and the mirror, and the angle-frequency (of the driving signal) response is shown in Fig. 7e as the red line. As Fig. 7e shows, around the mechanical resonant frequency of 990 Hz, the mirror doesn't vibrate at all. As the frequency increased, the oscillation starts at 1968 Hz, which is about twice the resonant frequency. The parametric resonant angle is ±1.87°.

To compare the angle of two resonant modes, the same 50-V square wave was also applied between actuator I and the mirror. The black line in Fig. 7e is the measured angle. An angle of ±2.05° at 983 Hz was achieved, that is, slightly higher than the angle under parametric resonance. This shows that the amplitude of harmonic resonance of the same device is not smaller than that of parametric resonance.

Because we use non-planar comb instead of coplanar comb, the mirror can also realize a static torsion. A DC voltage was applied between actuator I, IV and the scanner. Due to the symmetry, the driving voltage can only provide a static torque around y-axis. The scanner quasistatically deflected and a maximum optical angle of ±0.5° was achieved, as shown in Fig. 7d.

**X axis (fast axis) testing.** The harmonic and parametric resonance around x-axis is also tested. Figure 7b shows the frozen profile of the oscillation around the x-axis. A 100-V square wave was applied between actuator I and the mirror to excite the harmonic resonance. As for the parametric resonance, actuators I and IV is used due to they are located on both sides of x-axis. Thanks to the limitation of the turn-on voltage, a 115-V square wave was applied between actuators I, IV and the mirror. Figure 7f shows the scanning angle of the two resonances. The harmonic resonance has a resonant frequency of 5.96 kHz and a resonant angle of ±0.62°, so the measured frequency is very close to the simulated result. The parametric resonance occurs under a driving frequency of 11.91 kHz and reach a resonant angle of ±0.53°. Although the driving voltage of the harmonic oscillation is smaller, its angle is larger.

**Two axis testing.** In order to demonstrate the two-dimensional scanning ability of the proposed device, the slow axis and fast axis were driven simultaneously. The frequencies of the two axes are different, hence the profilometer cannot acquire a relative static surface profile. Actuator I was applied a square wave with 50 V voltage and 994 Hz frequency, actuator II was applied a square wave with 100 V voltage and 5.93 kHz frequency, and the mirror was grounded. The scanning pattern is shown in Fig. 8. Due to the mechanical filtering effect, the movement of the scanner is the superposition of two movements when the two actuators are driven separately. As mentioned earlier, one actuator with a superimposed signal of 50V square wave at 994 Hz and 100V square wave at 5.93 kHz is unavailable.

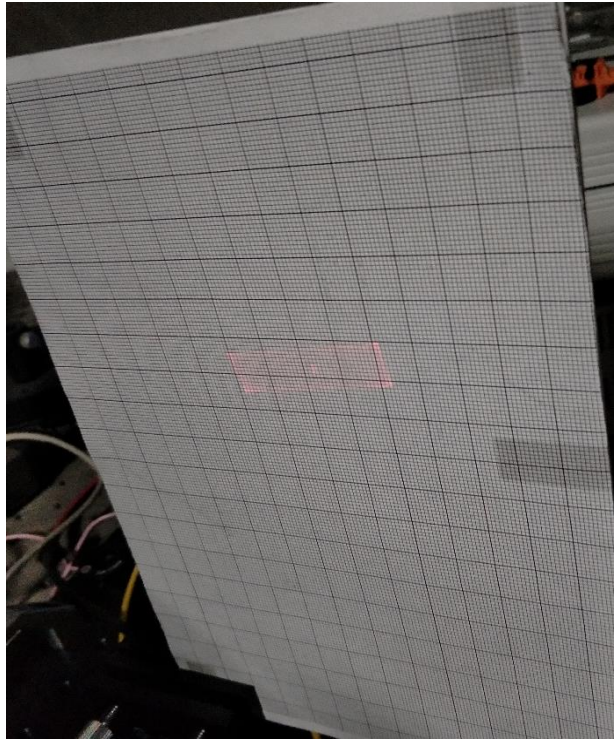

Fig. 8 Two-dimensional scanning.

## Discussion

A two-dimensional scanning mirror based on four identical electrostatic non-planar comb drive actuators is proposed. Its design, simulation, fabrication, and measurement are detailed. Through the breaking of the symmetry, harmonic resonance instead of parametric resonance is realized. It has the advantages of lower frequency of driving signal and larger scanning angle. Multiple scanning modes, including quasistatic scanning, harmonic resonant scanning, and parametric resonant scanning around two orthogonal axis, can be produced through different combinations of actuators. Based on mechanical filter effect, by applying different driving signal to different actuators, two-dimensional scanning without special or difficult insulation is realized through a cost-effective Multi Project Wafer (MPW) process.

## Materials and methods

**Finite elements analysis.** The FEM results of Fig. 3 were obtained using commercial finite element software COMSOL Multiphysics. The Solid Mechanics physics interface, Eigenfrequency study, and Eigenvalue Solver were used. The material was chosen as isotropic single-crystal silicon in the material library of the software, and Young's modulus, Poisson's ratio, and density are 170 GPa, 0.28, and 2.329 kg/m$^3$, respectively. Geometric nonlinearity was included in the simulation to model the structure more precisely.

**Sample fabrication.** The scanner was fabricated by SOIMUMPs, which has the general features of a standard SOI bulk micromachining process. The silicon layer is used as the structural material and its thickness is 25 μm. The Pad Metal and Blanket Metal are applied as electronic connection and self-assembly material and their thicknesses are 520 nm and 650 nm. The thicknesses of Oxide and Substrate are 2 μm and 400 μm, respectively.

## Acknowledgements

We acknowledge support from the Through Train Project for Ph. D. of Chongqing (CSTB2022BSXM-JSX0004), the Fundamental Research Funds for the Central Universities (2021CDJQY-041), and the CAS Light of West China Program.

## Conflict of interest

The authors declare no competing interests.

## Author contributions

W.W. conceived the original idea, carried out the experiment and supervised the project. W.W. and W.M. contributed to the numerical calculation and wrote the manuscript. All the authors have discussed the results.